\documentclass[manuscript,floatfix]{aastex63}
\usepackage{amsmath}
\usepackage{comment}
\usepackage[normalem]{ulem}

\let\vec\mathbf 

\begin{document}
\title{On the Conservation of Turbulence Energy in Turbulence Transport Models}
\author{B.-B. Wang}
\affiliation{Center for Space Plasma and Aeronomic Research (CSPAR), University of Alabama in
Huntsville, Huntsville, AL 35899, USA}
\author{G.P. Zank}
\affiliation{Center for Space Plasma and Aeronomic Research (CSPAR), University of Alabama in
Huntsville, Huntsville, AL 35899, USA}
\affiliation{Department of Space Science, University of Alabama in Huntsville, Huntsville, AL 35899,
USA}
\author{L. Adhikari}
\affiliation{Center for Space Plasma and Aeronomic Research (CSPAR), University of Alabama in
Huntsville, Huntsville, AL 35899, USA}
\affiliation{Department of Space Science, University of Alabama in Huntsville, Huntsville, AL 35899,
USA}
\author{L.-L. Zhao}
\affiliation{Center for Space Plasma and Aeronomic Research (CSPAR), University of Alabama in
Huntsville, Huntsville, AL 35899, USA}
\affiliation{Department of Space Science, University of Alabama in Huntsville, Huntsville, AL 35899,
USA}

\begin{abstract}
    Zank et al. developed models describing the transport of low frequency incompressible and nearly incompressible turbulence in inhomogeneous flows. The formalism was based on expressing the fluctuating variables in terms of the Els\"assar variables and then taking ``moments'' subject to various closure hypotheses. The turbulence transport models are different according to whether the plasma beta regime is large or of order 1. Here, we show explicitly that the two sets of turbulence transport models admit a conservation representation that resembles the well-known WKB transport equation for Alfv\'en wave energy density after introducing appropriate definitions of the ``pressure'' associated with the turbulent fluctuations. This includes introducing a distinct turbulent pressure tensor for 3D incompressible turbulence (the large plasma beta limit) and pressure tensors for quasi-2D and slab turbulence (the plasma beta order 1 regimes) that generalize the form of the WKB pressure tensor. Various limits of the different turbulent pressure tensors are discussed. However, the analogy between the conservation form of the turbulence transport models and the WKB model is not close for multiple reasons, including that the turbulence models express fully nonlinear physical processes unlike the strictly linear WKB description. The analysis presented here serves both as a check on the validity and correctness of the turbulence transport models and provides greater transparency of the energy dissipation term and the ``turbulent pressure'' in our models, which is important for many practical applications.
\end{abstract}

\section{Introduction}

The transport of incompressible MHD turbulence in inhomogeneous flows is a fundamentally important problem for both space physics and astrophysics, particularly in the context of the transport and acceleration of energetic particles such as solar energetic particles and galactic cosmic rays. Historically, the transport of incompressible fluctuations in a large-scale inhomogeneous flow has been modelled on the basis of a linear Alfv\'en wave description, colloquially known as the WKB model \citep{Parker_1965,Hollweg_1973}, and has been popular due to its tractability and simplicity. Being a linearized wave description, the leading-order 
 WKB model describes non-interacting propagating waves and neglects possible mixing or coupling between propagating modes (although see the higher-order corrections discussed by \cite{Heinemann_Olbert_1980}). The need to incorporate turbulence effects explicitly was recognized in the 1990's with the development of transport models that departed from the assumption of linearized modes and incorporated the mode mixing and nonlinear dissipation via the energy cascade through the inertial range 
\citep{Zhou-Matthaeus1989, Marsch_Tu_1990, Zhou_Matthaeus_1990, Zhou_Matthaeus_1990a, Matthaeus_etal_1994, Zank_etal_1996, Breech_etal_2008, Oughton_etal_2011, Zank_etal_2012, Zank_etal_2017}. Some discussion was presented by \cite{Matthaeus_etal_1994b} about the connection of the earlier turbulence models to the WKB description. However, the connection of the simpler WKB description to the much more elaborate turbulence transport models of \cite{Zank_etal_1996, Breech_etal_2008, Oughton_etal_2011,
Zank_etal_2012, Zank_etal_2017, Adhikari_etal_2017} has not been established. This paper addresses the connection between detailed turbulence transport models, their conservation form, and their relation to WKB models.

The application of the plasma beta large and plasma beta order one or less turbulence transport models to the solar wind, \citet{Zank_etal_2012} and \citet{Zank_etal_2017},  has shown good agreement with a large variety of observations \citep{Adhikari_etal_2015, Shiota_etal_2017, Adhikari_etal_2017, Zank_etal_2018, Adhikari_etal_2020, Adhikari_etal_2020a, Adhikari_etal_2020b, Zhao_etal_2020}. 
The derivation of the turbulence models is based on a two-scale separation method, which is a common approach to obtain turbulence transport models \citep{Zhou_Matthaeus_1989, Zhou_Matthaeus_1990a, Zhou_Matthaeus_1990, Zank_etal_1996, Zank_2014,Zank_etal_2012, Zank_etal_2017}.
Since the fluctuations are well separated from the scale associated with the large-scale inhomogeneities,
the MHD variables can be decomposed into small-scale rapidly varying fluctuations and large-scale slowly varying mean values.
The fluctuations are random variables with zero mean but can have an arbitrarily large amplitude. 
By applying an ensemble average operator $\langle \dots \rangle$ to the MHD equations, i.e., the mass continuity, momentum, energy, and Faraday's induction equations, we can obtain a system of evolution equations for the mean fields that are coupled to the fluctuating fields. 
On subtracting the operator-averaged equations from the original MHD equations, we obtain the evolution equations for the fluctuating fields.
The fluctuating fields can be combined and expressed in terms of the fluctuating Els\"asser variables, $\vec z^{\pm} \equiv \vec u \pm \vec b /\sqrt{4\pi\rho}$, and 
 $\vec u$, $\vec b$ and $\rho$ are the fluctuating or turbulent velocity and magnetic field, and the mean plasma mass density, respectively.
The dynamical equations for $\partial \vec z^{\pm}/\partial t$ are the basis for constructing a turbulence transport model.
By computing the second-order moments of $\vec z^{\pm}$ through the dynamical equations describing the evolution of the Els\"asser variables,
we can derive systems of equations describing the evolution of the moments $\langle \vec {z^+}^2 \rangle$, $\langle \vec {z^-}^2 \rangle$, and $\langle \vec z^+ \cdot \vec z^- \rangle$.
Such one-point closure schemes are utilized to derive the dissipation terms and the corresponding evolution equations for the correlation lengths.
The nonlinear terms that arise in the evolution equations for both the turbulence quantities and the correlation functions are simplified by using a structural similarity hypothesis. The structural similarity hypothesis is an approximation (essentially a closure hypothesis) for the variance and covariance of the field components as a fraction of the variance of the total, or equivalently relating the off-diagonal elements of the variance or covariance tensor to the corresponding trace.
For more details, we refer to \citet{Zank_etal_2012, Zank_etal_2017}. 

\citet{Zank_etal_2012, Zank_etal_2017} derived a coupled system equations that describes the transport of the Els\"asser energy in backward propagating modes $\langle {\vec z^+}^2 \rangle$, forward propagating modes $\langle {\vec z^{-}}^2 \rangle$, the cross helicity $E_C \equiv \langle {\vec z^+}^2 \rangle - \langle {\vec z^-}^2 \rangle$, the residual energy $E_D \equiv \langle \vec z^+ \cdot \vec z^- \rangle$, and correlation lengths corresponding to forward propagating
modes $\lambda^+$, backward propagating modes $\lambda^-$, and the residual energy $\lambda_D$.
Under an additional set of simplifying assumptions, the large plasma beta model of \citet{Zank_etal_2012} can be further reduced to a single transport equation in the magnetic energy density as derived by \citet{Zank_etal_1996} (see also \citep{Adhikari_etal_2020c}), from which one can recover the well known WKB model after neglecting the dissipation and mixing terms \citep{Zank_etal_2012}.

The focus of this work is to present a conservation form of the two sets of turbulence transport equations that were derived in the beta large and beta order one regimes.  
This analysis serves both as a check on the validity and correctness of the transport models and provides greater transparency of the energy dissipation term and the ``turbulent pressure'' in our models, which is important for many practical applications.
The importance of the dissipation of turbulence is of course related to  the heating of gas or plasma in numerous space and astrophysical environments, especially in the heating of the solar corona and the acceleration of the solar wind \citep{Matthaeus_etal_1999a, Dmitruk_etal_2001, Oughton_etal_2001, Dmitruk_etal_2002, Cranmer_etal_2007, Chandran_Hollweg_2009, Verdini_etal_2010, Woolsey_Cranmer_2014, Ballegooijen_Asgari_2016, Zank_etal_2018a, Adhikari_etal_2020, Adhikari_etal_2020b}, and the heating of the extended heliosphere \citep{Matthaeus_etal_1999, Smith_etal_2001, Isenberg_etal_2003, Isenberg_2005, Adhikari_etal_2017,Montagud_Camps_etal_2018, Zank_etal_2018, Adhikari_etal_2020a}.
Besides the effects of turbulent dissipation in heating the thermal gas, turbulence can contribute to the dynamical behavior of a gas via its contribution to the total pressure. This has been of particular interest in the context of shock waves mediated by cosmic rays, where the pressure contributed by the turbulence excited by cosmic ray streaming decelerates the gas flow, thereby changeing the shock profile and modifying the accelerated or energetic particle spectrum \citep{McKenzie_Voelk_1982,Jones_1993,Ko_1995,Caprioli_etal_2009}.

\section{Transport of turbulent energy}
It is convenient to represent turbulence quantities by a set of one-point moments of the Els\"asser variables, 
\begin{gather}
    E_T \equiv \langle \vec u^2+ \vec b^2/(4\pi\rho) \rangle = \frac{\langle \vec {z^+}^2 \rangle + \langle \vec {z^-}^2 \rangle}{2} ; \label{eq:1} \\
    E_C \equiv 2 \langle \vec u \cdot \vec b/(\sqrt{4\pi\rho}) \rangle = \frac{\langle \vec {z^+}^2 \rangle - \langle \vec {z^-}^2 \rangle}{2} ;\label{eq:2}  \\
    E_D \equiv \langle \vec u^2- \vec b^2/(4\pi\rho) \rangle = \langle \vec {z^+} \cdot \vec z^-  \rangle , \label{eq:3}
\end{gather}
where $E_T$ is twice the total turbulent kinetic and magnetic energy per unit mass, $E_C$ is the cross helicity measuring the correlation between the fluctuating velocity and magnetic fields, and $E_D$ is the residual energy representing the difference between (twice) the turbulent kinetic and magnetic energies per unit mass. 
The normalized cross helicity and residual energy are defined as $\sigma_C=E_C/E_T$ and $\sigma_D=E_D/E_T$, respectively.
Our focus here is on the turbulence energy density $E_w$, which is defined as the sum of the turbulence kinetic and magnetic energy densities,
\begin{equation}
    E_w \equiv  \frac{\rho}{2} \langle \vec u^2 \rangle + \frac{\langle \vec b^2 \rangle}{8\pi} = \frac{\rho}{2} E_T.  \label{eq:4}
\end{equation}

\subsection{Transport of incompressible turbulence in the plasma beta large regime}\label{sec:2012}
Consider first the turbulence transport equations derived from the 3D incompressible MHD equations. As discussed in \cite{Zank_Matthaeus_1993}, the 3D incompressible MHD equations represent the leading order description of nearly incompressible MHD in the limit of large plasma beta, and can be derived from the Els\"asser variables representation introduced by \cite{Zhou_Matthaeus_1989, Zhou_Matthaeus_1990a, Zhou_Matthaeus_1990, Marsch_Tu_1989}. 
The 3D time dependent turbulence transport model is then given by \citep{Zank_etal_2012},
\begin{align}
\begin{split}
 &\frac{\partial E_{T}}{\partial t} +\vec{U} \cdot \nabla E_{T} +\nabla \cdot \vec{V}_{A} E_{C} -\vec{V}_{A} \cdot \nabla E_{C} +\nabla \cdot \vec{U}\left[\frac{E_{T}}{2} +\left( 2a-\frac{1}{2}\right) E_{D}\right] -2aE_{D}\vec{nn} :\nabla \vec{U} =\\
 &\qquad -\frac{( E_{T} +E_{C})( E_{T} -E_{C})^{1/2}}{\lambda ^{+}} -\frac{( E_{T} -E_{C})( E_{T} +E_{C})^{1/2}}{\lambda ^{-}}  ; \label{eq:2012}
\end{split}\\ 
\begin{split}
    &\frac{\partial E_{C}}{\partial t} +\vec{U} \cdot \nabla E_{C} +\frac{1}{2} \nabla \cdot \vec{U} E_{C} -\vec{V}_{A} \cdot \nabla E_{T} +\nabla \cdot \vec{V}_{A} E_{T} - \nabla \cdot \vec{V}_{A} E_{D} -2bE_{D}\vec{nn} :\nabla \vec{B}/\sqrt{4\pi\rho} \\
&\qquad  =-\frac{( E_{T} +E_{C})( E_{T} -E_{C})^{1/2}}{\lambda ^{+}} +\frac{( E_{T} -E_{C})( E_{T} +E_{C})^{1/2}}{\lambda ^{-}} ; 
\end{split}\\
\begin{split}
    &\frac{\partial E_{D}}{\partial t} +\vec{U} \cdot \nabla E_{D} +\frac{1}{2} \nabla \cdot \vec{U} E_{D} +\left( 2a-\frac{1}{2}\right) \nabla \cdot \vec{U} E_{T}  +\frac{E_{C}\vec{V}_{A} \cdot \nabla E_{T} 
     -E_{T}\vec{V}_{A} \cdot \nabla E_{C}}{E^{2}_{T} -E^{2}_{C}}  \\
     & \quad +\frac{1}{2} \nabla \cdot \vec{V}_{A} E_{C}
     -2\vec{nn} :( aE_{T} \nabla \vec{U} -bE_{C} \nabla \vec{B}/\sqrt{4\pi\rho}) =-E_D\frac{\vec{V}_A}{\lambda_D}+
    \frac{(E_T-E_C)( E_{T} +E_{C})^{1/2}}{\lambda ^{-}} \\
&\qquad +\frac{(E_T+E_C) (E_{T} -E_{C})^{1/2}}{\lambda ^{+}}; 
\end{split}\\
\begin{split}
    &\frac{\partial \lambda ^{\pm }}{\partial t} +(\vec{U} \mp \vec{V}_{A}) \cdot \nabla \lambda ^{\pm } +\frac{E_{D}}{E_{T} \pm E_{C}}\bigg[\left( a-\frac{1}{4}\right) \nabla \cdot \vec{U} \mp  \frac{1}{2} \nabla \cdot \vec{V}_{A} \mp b\vec{nn} :\nabla \vec{B}/\sqrt{4\pi\rho}\\ 
    &\qquad -a\vec{nn} :\nabla \vec{U}\bigg]\left( \lambda _{D} -2\lambda ^{\pm }\right) =2( E_{T} \mp E_{C})^{1/2};
\end{split}\\
\begin{split}
&\frac{\partial \lambda _{D}}{\partial t} +\vec{U} \cdot \nabla \lambda _{D} +\frac{2E_{T}}{E_{D}}\left[\left( a-\frac{1}{4}\right) \nabla \cdot \vec{U} -a\vec{nn} :\nabla \vec{U}\right]\left[\frac{( E_{T} +E_{C}) \lambda ^{+} +( E_{T} -E_{C}) \lambda ^{-}}{E_{T}} -\lambda _{D}\right]\\
    &\qquad -\frac{2E_{C}}{E_{D}}\left[-\frac{1}{2} \nabla \cdot \vec{V}_{A} -b\vec{nn} :\nabla \vec{B}/\sqrt{4\pi\rho}\right]  \left[\frac{( E_{T} +E_{C}) \lambda ^{+} -( E_{T} -E_{C}) \lambda ^{-}}{E_{C}} -\lambda _{D}\right] \\
    &\qquad +\frac{E_{C}\vec{V}_{A} \cdot \nabla E_{T} -E_{T}\vec{V}_{A} \cdot \nabla E_{C}}{E_{D}\sqrt{E^{2}_{T} -E^{2}_{C}}}
 \left( 2\sqrt{\lambda ^{+} \lambda ^{-}} -\lambda _{D}\right)  +\frac{\sqrt{E^{2}_{T} -E^{2}_{C}}}{E_{D}}\bigg[\left(\frac{\lambda^{+}}{\lambda ^{-}}\right)^{1/2}\vec{V}_{A} \cdot \nabla \lambda ^{-} - \\
    &\qquad \left(\frac{\lambda ^{-}}{\lambda ^{+}}\right)^{1/2}\vec{V}_{A} \cdot \nabla \lambda ^{+}\bigg]
 =V_{A} -\left[\frac{( E_{T} +E_{C})( E_{T} -E_{C})^{1/2}}{\lambda ^{+}} +\frac{( E_{T} -E_{C})( E_{T} +E_{C})^{1/2}}{\lambda ^{-}}\right]\frac{\lambda _{D}}{E_{D}} ,
\end{split}
\end{align}
where $\vec{U}$ is the large-scale fluid velocity, $\vec{V}_A$ the large-scale Alfv\'en velocity, $\lambda^\pm$ the correlation length for forward/backward propagating modes, and $\vec{n}$ corresponds to a specified direction for axisymmetric turbulence (typically the imposed mean magnetic field direction).
The parameters 
$a$ and $b$ are structural similarity parameters and their origin in the context of the transport model above is a little subtle \citep{Zank_etal_2012}. Specifically, $a$ is a closure that relates the off-diagonal elements of the second-order tensors $\langle z_i z_j \rangle$ (where we deliberately leave the superscripts $\pm$ off to indicate generality) to the trace through $a \; (\mbox{or} \; b) \langle z^2 \rangle$. Since $\langle z_i z_j \rangle$ occurs in conjunction with the gradient of either the large-scale flow velocity $\vec{U}$ or the Alfv\'en velocity $\vec V_A$, $a$ is associated with gradients in the large-scale flow ${\bf U}$ whereas the structural similarity parameter $b$ is associated specifically with gradients in ${\bf V}_A$.
The choice of $a=b=1/2$, or $a=1/3$ corresponds to either the 2D or the 3D mixing tensor in the \citet{Matthaeus_etal_1994, Zank_etal_1996} turbulence transport model.
For 3D isotropic turbulence, the axis-symmetric direction vector $\vec{n}$ should be a zero vector and disappears together with parameter $b$. 

In deriving the energy conservation equation, we need some essential vector and tensor relations,
\begin{gather}
\nabla \cdot ( \alpha \vec A) = \alpha \nabla \cdot \vec A + \nabla \alpha \cdot \vec A 	\label{eq:divv}; \\
\vec T: \nabla \vec A = \nabla \vec A: \vec T \label{eq:scalar}; \\
\nabla \cdot \vec A = \nabla \vec A : \vec I \label{eq:unit}; \\ 
\nabla (\alpha \vec T) = \alpha \nabla \cdot \vec T + \nabla \alpha \cdot \vec T  \label{eq:divT};\\
\nabla \cdot (\vec A \cdot \vec T) = \vec A \cdot \nabla \cdot \vec T + \vec T : \nabla \vec A \label{eq:divvT},
\end{gather}
where $\alpha$ is a scalar, $\vec A$ is a vector, $\vec T$ is a tensor, and $\vec I$ is a identity tensor.

On neglecting the dissipation and source terms in Equation (\ref{eq:2012}), the transport equation for $E_T$ can be written as,
\begin{equation}
\frac{\partial E_{T}}{\partial t} +\vec{U} \cdot \nabla E_{T} -\vec{V}_{A} \cdot \nabla E_{C} +\nabla \cdot \vec{V}_{A} E_{C} +\nabla \cdot \vec{U}\left[\frac{E_{T}}{2} +\left( 2a-\frac{1}{2}\right) E_{D}\right] -2aE_{D}\vec{nn} :\nabla \vec{U} =0. \label{eq:E_T0}
\end{equation}
Using of Equation (\ref{eq:divv}) on the second and third terms of (\ref{eq:E_T0}) and adding the fourth term yields 
\begin{equation}
\vec{U} \cdot \nabla E_{T} -\vec{V}_{A} \cdot \nabla E_{C} +\nabla \cdot \vec{V}_{A} E_{C} =\nabla \cdot (\vec{U} E_{T} -\vec{V}_{A} E_{C}) -E_{T} \nabla \cdot \vec{U} +2E_{C} \nabla \cdot \vec{V}_{A} .
\end{equation}
By using Equations (\ref{eq:unit}) and (\ref{eq:scalar}), the fifth and the last terms on the left hand side of Equation (\ref{eq:E_T0}) become 
\[
\nabla \cdot \vec{U}\left[\frac{E_{T}}{2} +\left( 2a-\frac{1}{2}\right) E_{D}\right] -2aE_{D}\vec{nn} :\nabla \vec{U} = \nabla \vec U : \frac{2}{\rho} \vec P_w ,
\]
where we have introduced the turbulence pressure tensor $\vec P_w$,
\begin{equation} 
\vec P_w \equiv \frac{\rho}{2} \left[ \left( \frac{E_{T}}{2} +\left( 2a-\frac{1}{2}\right) E_{D} \right) \vec I - 2aE_D \vec n \vec n \right] . \label{eq:Pw}
\end{equation}
Equation (\ref{eq:E_T0}) can therefore be expressed as
\begin{equation}
    \frac{\partial E_{T}}{\partial t} +\nabla \cdot (\vec{U} E_{T} -\vec{V}_{A} E_{C}) +\nabla \vec{U} :\left(\frac{2}{\rho}\vec{P}_{w}\right) -E_{T} \nabla \cdot \vec{U} +2E_{C} \nabla \cdot \vec{V}_{A} = 0 . \label{eq:15a}
\end{equation}
After substituting $E_C=\sigma_c E_T$, $E_D=\sigma_D E_T$, and $E_T=\rho/2 E_w$ into Equation (\ref{eq:15a}), and multiplying by $\rho/2$, we obtain
\begin{eqnarray}
\frac{\partial E_{w}}{\partial t} +\nabla \cdot [(\vec{U} -\vec{V}_{A} \sigma _{C}) E_{w}] +\nabla \vec{U} :\vec{{P}}_{w} +\rho E_{w}\frac{\partial }{\partial t}\left(\frac{1}{\rho } \right)\nonumber \\ 
\mbox{}    +\rho (\vec{U} -\vec{V}_{A} \sigma _{c}) \cdot \nabla \left(\frac{1}{\rho }\right) 
-\frac{E_{w}}{\rho } \rho \nabla \cdot \vec{U} +2\sigma _{c} E_{w} \nabla \cdot \vec{V}_{A} =0.  \label{eq:15b} 
\end{eqnarray}

Since $\nabla \cdot \vec V_A = -\vec V_A \cdot \nabla \rho /(2\rho)$, the last four terms can be eliminated as follows:
\begin{eqnarray}
    \rho E_{w}\frac{\partial }{\partial t}\left(\frac{1}{\rho }\right) &+& \rho E_{w}(\vec{U} -\vec{V}_{A} \sigma _{c}) \cdot \nabla \left(\frac{1}{\rho }\right) -\frac{E_{w}}{\rho } \rho \nabla \cdot \vec{U} +2\sigma _{c} E_{w} \nabla \cdot \vec{V}_{A} \nonumber \\
     \qquad  &=& -\frac{E_{w}}{\rho }\left(\frac{\partial \rho }{\partial t} +\vec{U} \cdot \nabla \rho -\sigma _{c}\vec{V}_{A} \cdot \nabla \rho +\rho \nabla \cdot \vec{U} +\sigma _{c}\vec{V}_{A} \cdot \nabla \rho \right) \nonumber  \\
       &=& -\frac{E_{w}}{\rho }\left[\frac{\partial \rho }{\partial t} +\nabla \cdot ( \rho \vec{U})\right] \nonumber \\
    &=& 0, \label{eq:20} 
\end{eqnarray}
thanks to conservation of mass.
Finally, using (\ref{eq:divvT}), we can express the transport equation for the total turbulence energy in a conservative form resembling that of a WKB model,
\begin{equation}
\frac{\partial E_{w}}{\partial t} +\nabla \cdot \left[ (\vec{U} -\sigma_{C}\vec{V}_{A} ) E_{w} +\vec{U\cdot \vec{P}}_{w} \right] =\vec{U} \cdot \nabla \cdot {\vec{P}}_{w}. \label{eq:16}
\end{equation}
The term in square brackets is the energy density flux vector, which is the amount of turbulence energy passing in unit time through a unit area perpendicular to the direction of the velocity \citep{Landau_Lifshitz_1987}.
Within the square brackets, the first term is the energy transported through the unit surface area in unit time, and the second term is the work done by the turbulent ``pressure'' force on the plasma within the surface. 
The right hand term is the rate of work of the turbulence pressure gradient on the background plasma flow. For the present, we remind the reader that we have neglected the dissipation term in deriving Equation (\ref{eq:16}) -- this term is given below.

The turbulence propagation velocity is the energy-averaged Alfv\'en velocity,
\begin{equation}
    -\sigma_C \vec V_A=-\frac{\langle \vec {z^+}^2 \rangle - \langle \vec {z^-}^2 \rangle }{ \langle \vec {z^+}^2 \rangle + \langle \vec {z^-}^2 \rangle} \vec V_A = \frac{ \langle \vec {z^-}^2 \rangle }{ \langle \vec{z^+}^2 \rangle +\langle \vec{z^-}^2 \rangle}\vec V_A -\frac{\langle \vec {z^+}^2 \rangle}{\langle \vec{z^+}^2 \rangle + \langle \vec{z^-}^2 \rangle}\vec V_A, \label{eq:22}
\end{equation}
which resembles the mean local velocity of Alfv\'enic turbulence. 
Since the turbulence consists of structures that move in all directions, the mean local velocity of the Alfv\'en turbulence is the energy-averaged Alfv\'en velocity weighted by the ratio of the forward or backward wave energy to the total energy \citep{Bell_Lucek_2001}.

We can express the turbulent pressure tensor in terms of the turbulence energy and the fluctuating fields as 
\begin{align} 
    \vec P_w &= \left[ \frac{E_{w}}{2} +\left( 2a-\frac{1}{2}\right) \sigma_{D}E_w \right] \vec I - 2a\sigma_DE_w \vec n \vec n  \nonumber \\ 
       & = \left[ a\rho \langle \vec u^2 \rangle + (1-2a) \left( \frac{\langle \vec b^2 \rangle}{8\pi} \right)\right] \vec I -a\left(\rho \langle \vec u^2 \rangle - \frac{\langle \vec b^2 \rangle}{4\pi}\right) \vec{nn} .   \label{eq:23}
\end{align}
It is worth noting that if $a>1/2$, it is possible for the isotropic part of the pressure tensor to be negative.
For Alfv\'en-like turbulence with $E_D=0=\sigma_D$, the turbulence pressure is the familiar isotropic Alfv\'en wave pressure $\vec b^2/(8\pi)\vec{I}$.
For 3D isotropic turbulence, $a=1/3$ and $\vec{n}=0$, the reduced turbulence pressure tensor is
\begin{equation}
    \vec P_w = \frac{\rho}{4}(E_T+\frac{1}{3}E_D) = \frac{\rho \langle \vec u^2 \rangle }{3}+\frac{1}{3}\left( \frac{ \langle \vec b^2 \rangle}{8\pi} \right).
\end{equation}
Note that the turbulence pressure, including the ``ram pressure'' (i.e., the kinetic or fluctuating Reynold's pressure) and the fluctuating magnetic stress, is \citep{McKee_Zweibel_1995}
\begin{equation}
    \vec P =  \rho \langle \vec u \vec u \rangle + \frac{\langle \vec b^2 \rangle }{8\pi} \vec{I} - \frac{ \langle \vec b\vec b \rangle}{4\pi}. \label{eq:wave_pressure}
\end{equation}
In the case of 3D isotropic turbulence, the local average fluctuating ram pressure $\rho \langle u_iu_j \rangle =\rho \langle u_i^2 \rangle$ is $\rho \langle u^2 \rangle/3$ since the average of $u_iu_j=0$ for $i \neq j$.
This is true too for the fluctuating magnetic stress, and is given by ${\langle \vec b^2 \rangle}/(8\pi)-({\langle \vec b^2 \rangle}/(4\pi))/3 = (\langle \vec b^2 \rangle/(8\pi))/3$. 

For turbulence that is axisymmetric with respect to the directional vector $\vec n$, and has $a=1/2$, the turbulence pressure tensor is given by 
\begin{equation}
    \vec P_w = \frac{\rho \langle u^2 \rangle}{2}(\vec I - \vec n \vec n) + \frac{\langle b^2 \rangle}{8\pi} \vec n \vec n. \label{eq:pressue_0.5}
\end{equation}
This results from our structural similarity assumptions for 2D turbulence \citep{Zank_etal_2012},
\begin{equation}
   \langle \vec u \vec u \rangle= \frac{1}{2} \langle \vec u^2 \rangle \vec I - \frac{1}{2}\langle \vec u^2 \rangle \vec n \vec n ; \qquad 
   \langle \vec b \vec b \rangle= \frac{1}{2}\langle \vec b^2  \rangle \vec I - \frac{1}{2}\langle \vec b^2 \rangle \vec n \vec n ,  \label{eq:27} 
\end{equation}
which, when substituted into the turbulence pressure equation (\ref{eq:wave_pressure}), yields Equation (\ref{eq:pressue_0.5}).

The dissipation of turbulence energy $E_{diss}$ is easily found to be given by \citep{Zank_etal_2012}
\begin{equation}
    E_{diss} = -\frac{\rho}{2} \left[ \frac{( E_{T} +E_{C})( E_{T} -E_{C})^{1/2}}{\lambda ^{+}} -\frac{( E_{T} -E_{C})( E_{T} +E_{C})^{1/2}}{\lambda ^{-}} \right] . \label{eq:28}
\end{equation}
The complete energy transport equation, including the dissipation term, is therefore given by 
\begin{align}
    \frac{\partial E_{w}}{\partial t} +\nabla \cdot [(\vec{U} -\vec{V}_{A} \sigma _{C}) E_{w} +\vec{U\cdot \vec{P}}_{w}] &=\vec{U} \cdot \nabla \cdot {\vec{P}}_{w} - \frac{\rho}{2} \left[ \frac{( E_{T} +E_{C})( E_{T} -E_{C})^{1/2}}{\lambda ^{+}}  \right. \nonumber \\ 
& - \left. \frac{( E_{T} -E_{C})( E_{T} +E_{C})^{1/2}}{\lambda ^{-}} \right] .  \label{eq:29}
\end{align}

\subsection{Transport of nearly incompressible 2D turbulence in an inhomogeneous $\beta \sim 1$  plasma} 
From the perspective of nearly incompressible MHD, the incompressible MHD description is valid only for a plasma beta regime much large than one; for a plasma beta order one or less, the turbulence is a superposition of a dominant 2D incompressible component and a minority slab component \citep{Zank_Matthaeus_1992, Zank_Matthaeus_1993, Zank_etal_2017, Zank_etal_2020}. The equations governing the evolution of 2D incompressible turbulence in the plasma beta order one limit are  \citep{Zank_etal_2017} 
\begin{flalign}
\begin{split}
    & \frac{\partial E^{\infty }_{T}}{\partial t} +\vec{U} \cdot \nabla E^{\infty }_{T} +\nabla \cdot \vec{U}\left[\frac{E^{\infty }_{T}}{2} +\left( 2a-\frac{1}{2}\right) E^{\infty }_{D}\right] =\frac{\vec{n} \cdot \nabla \rho }{4\rho } \bigg[ \left( E^{\infty }_{T} +E^{\infty }_{C}\right)^{3/2} +\left( E^{\infty }_{T} -E^{\infty }_{C}\right)^{3/2} \\
    & \qquad -  E^{\infty }_{D}\left( E^{\infty }_{T} +E^{\infty }_{C}\right)^{1/2} -E^{\infty }_{D}\left( E^{\infty }_{T} -E^{\infty }_{C}\right)^{1/2} \bigg] -\frac{\left( E^{\infty }_{T} +E^{\infty }_{C}\right)\left( E^{\infty }_{T} -E^{\infty }_{C}\right)^{1/2}}{\lambda ^{+}_{\perp }}  \\
    & \qquad - \frac{\left( E^{\infty }_{T} -E^{\infty }_{C}\right)\left( E^{\infty }_{T} +E^{\infty }_{C}\right)^{1/2}}{\lambda ^{-}_{\perp }} ; \qquad \label{eq:2D} 
\end{split} \\
\begin{split}
    & \frac{\partial E^{\infty }_{C}}{\partial t} +\vec{U} \cdot \nabla E^{\infty }_{C} +\frac{\nabla \cdot \vec{U}}{2} E^{\infty }_{C} = \frac{\vec{n} \cdot \nabla \rho }{4\rho } \bigg[ \left( E^{\infty }_{T} +E^{\infty }_{C}\right)^{3/2} -\left( E^{\infty }_{T} -E^{\infty }_{C}\right)^{3/2} \\
    & \qquad +E^{\infty }_{D}\left( E^{\infty }_{T} -E^{\infty }_{C}\right)^{1/2} -  E^{\infty }_{D}\left( E^{\infty }_{T} +E^{\infty }_{C}\right)^{1/2} \bigg] -\frac{\left( E^{\infty }_{T} +E^{\infty }_{C}\right)\left( E^{\infty }_{T} -E^{\infty }_{C}\right)^{1/2}}{\lambda ^{+}_{\perp }} \\
    & \qquad +\frac{\left( E^{\infty }_{T} -E^{\infty }_{C}\right)\left( E^{\infty }_{T} +E^{\infty }_{C}\right)^{1/2}}{\lambda ^{-}_{\perp }}  ; \qquad \label{eq:31}  
\end{split}\\
\begin{split}
    & \frac{\partial E^{\infty }_{D}}{\partial t} +\vec{U} \cdot \nabla E^{\infty }_{D} +\nabla \cdot \vec{U}\left[\frac{E^{\infty }_{D}}{2} +\left( 2a-\frac{1}{2}\right) E^{\infty }_{T}\right] =\frac{\vec{n} \cdot \nabla \rho }{4\rho } \bigg[ E^{\infty }_{D}\left( E^{\infty }_{T} +E^{\infty }_{C}\right)^{1/2}  \\ 
    & \qquad+E^{\infty }_{D}\left( E^{\infty }_{T} -E^{\infty }_{C}\right)^{1/2}  -  \left( E^{\infty }_{T} +E^{\infty }_{C}\right)\left( E^{\infty }_{T} -E^{\infty }_{C}\right)^{1/2} -\left( E^{\infty }_{T} -E^{\infty }_{C}\right)\left( E^{\infty }_{T} +E^{\infty }_{C}\right)^{1/2} \bigg] \\
    & \qquad -E^{\infty }_{D} \left[ \frac{\left( E^{\infty }_{T} -E^{\infty }_{C}\right)^{1/2}}{\lambda ^{+}_{\perp }} 
   +  \frac{\left( E^{\infty }_{T} +E^{\infty }_{C}\right)^{1/2}}{\lambda ^{-}_{\perp }} \right]  ; \qquad  \label{eq:32} 
\end{split}\\
\begin{split}
    & \frac{\partial L^{\pm }_{\infty }}{\partial t} +\vec{U} \cdot \nabla L^{\pm }_{\infty } +\nabla \cdot \vec{U}\left[ \frac{L^{\pm }_{\infty }}{2} +\left( a-\frac{1}{4}\right) L^{\infty }_{D} \right] +\frac{\vec{n} \cdot \nabla \rho }{4\rho }\left( E^{\infty }_{T} \pm E^{\infty }_{C}\right)^{1/2}\left( L^{\infty }_{D} -2L^{\pm }_{\infty }\right) \\
    & \qquad = 0 ; \qquad \label{eq:33} 
\end{split}\\
\begin{split}
    & \frac{\partial L^{\infty }_{D}}{\partial t} +\vec{U} \cdot \nabla L^{\infty }_{D} +\nabla \cdot \vec{U}\left[ \frac{L^{\infty }_{D}}{2} +\left( 2a-\frac{1}{2}\right)\left( L^{+}_{\infty } +L^{-}_{\infty }\right) \right] -\frac{\vec{n} \cdot \nabla \rho }{4\rho }\bigg[ L^{\infty }_{D}\left( E^{\infty }_{T} +E^{\infty }_{C}\right)^{1/2} \\
    & \qquad + L^{\infty }_{D}\left( E^{\infty }_{T} -E^{\infty }_{C}\right)^{1/2} -2L^{+}_{\infty }\left( E^{\infty }_{T} -E^{\infty }_{C}\right)^{1/2} -2L^{-}_{\infty }\left( E^{\infty }_{T} +E^{\infty }_{C}\right)^{1/2}\bigg] \\
    & \qquad = 0,  \label{eq:34} 
\end{split}
\end{flalign}
where the subscript ``$\infty$'' denotes MHD variables that satisfy the incompressible equations, and ``$*$'' denotes the higher-order corrections. 
We assume that the structural similarity parameter for the fluctuating velocity and magnetic fields are the same and denoted by $a$. 
Note that in this (and next) subsection $\vec{n}$ is averaged direction vector of $z^{\infty\pm}$ in the local coordinate system along the large-scale magnetic field.

$L_{\infty}^{\pm}$ and $\lambda_{\perp}^{\pm}$ are the correlation function and correlation length corresponding to the forward/backward propagating modes, respectively, and 
$L_D^{\infty}$ and $\lambda_D^{\infty}$ are the correlation function and correlation length corresponding to the residual energy.
The correlation functions and correlation lengths are related by
\begin{eqnarray}
    L_{\infty}^{\pm} = \int\langle \vec z^{\infty \pm} \cdot \vec z^{\infty \pm'}\rangle dr = \langle {z^{\infty \pm}}^2 \rangle \lambda_{\perp}^{\pm} ; \label{eq"35} \\
L_{D}^{\infty} = \int\langle \vec z^{\infty +} \cdot \vec z^{\infty -'} + \vec z^{\infty +'} \cdot
    \vec z^{\infty -} \rangle dr = E_D^{\infty} \lambda_{D}^{\infty}, \label{eq:36}
\end{eqnarray}
where $\vec z^{\infty \pm '}\equiv \vec z^{\infty \pm}(\vec x + \vec r)$ indicates the lagged Els\"asser variable at location $\vec r$ from $\vec x$.

Were we to neglect the terms containing $V_A$ and $\vec n \vec n :\nabla \vec U$ on the left hand side of Equation (\ref{eq:2012}), we would obtain the same as the left-hand-side of Equation (\ref{eq:2D}). Thus, the conservation form of the evolution equation for the 2D turbulence energy is given immediately by 
\begin{align}
\begin{split}
    &\frac{\partial E_w^{\infty} }{\partial t} + \nabla \cdot \left( {\bf U} E_w^{\infty} + {\bf U} \cdot {\bf P}^{\infty}_w \right)  = {\bf U} \cdot \nabla \cdot {\bf P}^{\infty}_w + \frac{ {\bf n} \cdot \nabla \rho }{8} \bigg[ \left( E_T^{\infty} + E_C^{\infty} \right)^{3/2} + \left( E_T^{\infty} - E_C^{\infty} \right)^{3/2}   \\ 
& \qquad -  E^{\infty }_{D}\left( E^{\infty }_{T} +E^{\infty }_{C}\right)^{1/2} -E^{\infty }_{D}\left( E^{\infty }_{T} -E^{\infty }_{C}\right)^{1/2} \bigg] + \frac{\rho}{2} \bigg[ -\frac{\left( E^{\infty }_{T} +E^{\infty }_{C}\right)\left( E^{\infty }_{T} -E^{\infty }_{C}\right)^{1/2}}{\lambda ^{+}_{\perp }}   \\
& \qquad - \frac{\left( E^{\infty }_{T} -E^{\infty }_{C}\right)\left( E^{\infty }_{T} +E^{\infty }_{C}\right)^{1/2}}{\lambda ^{-}_{\perp }}\bigg]  \label{eq:37}
\end{split} 
\end{align}
The turbulence pressure tensor is now defined as
\begin{equation}
    \vec{P}_{w}^{\infty} \equiv  \left[ \frac{E_{w}^{\infty}}{2} +\left( 2a-\frac{1}{2}\right) \sigma_{D} E_{w}^{\infty}\right]\vec{I} =  \left[a\rho \langle \vec {u^\infty}^2 \rangle + (1-2a)\left (\frac{\langle \vec {b^\infty}^2 \rangle}{8\pi} \right)\right] \vec I, \label{eq:38} 
\end{equation}
where $u^\infty$ and $b^\infty$ are 2D turbulent velocity and magnetic field respectively, which was first noted by \citet{leRoux_etal_2018}.
The properties are similar to that for the incompressible turbulence case discussed in subsection \ref{sec:2012}.

\subsection{Transport of nearly incompressible slab turbulence in an inhomogeneous $\beta \sim \,1$ plasma}
The transport equations that describe the evolution of slab turbulence expressed in terms of the nearly incompressible corrections to the incompressible MHD variables are given by  \citep{Zank_etal_2017},
\begin{align}
\begin{split}
    & \frac{\partial E^{*}_{T}}{\partial t} +\vec{U} \cdot \nabla E^{*}_{T}+\nabla \cdot \vec{V}_A E_C^* -\vec{V}_{A} \cdot \nabla E^{*}_{C}  +\nabla \cdot \vec{U}\left[\frac{E^{*}_{T}}{2} +\left( 2b-\frac{1}{2}\right) E^{*}_{D}\right] -2bE^{*}_{D}\vec{ss} :\nabla \vec{U}  \\
    & \qquad   =\frac{\vec{n} \cdot \nabla \rho }{4\rho }\bigg[\left( E^{*}_{T} -E^{*}_{C}\right)(E_T^\infty-E_C^\infty)^{1/2} +\left( E^{*}_{T} +E^{*}_{C}\right)(E_T^\infty+E_C^\infty)^{1/2}  \\
    & \qquad - E_D^*\left(\sqrt{E_T^\infty+E_C^\infty}+\sqrt{E_T^\infty-E_C^\infty} \right)\bigg]  -\frac{\left( E^{\infty }_{T} -E^{\infty }_{C}\right)^{1/2}\left( E^{*}_{T} +E^{*}_{C}\right)}{\lambda ^{+}_{\perp }} \\
    & \qquad -\frac{\left( E^{\infty }_{T} +E^{\infty }_{C}\right)^{1/2}\left( E^{*}_{T} -E^{*}_{C}\right)}{\lambda ^{-}_{\perp }} ;\quad \label{eq:slab} 
\end{split} \\ 
\begin{split}
    & \frac{\partial E^{*}_{C}}{\partial t} +\vec{U} \cdot \nabla E^{*}_{C} -\vec{V}_{A} \cdot \nabla E^{*}_{T} +\frac{\nabla \cdot \vec{U}}{2} E^{*}_{C} + \nabla \cdot \vec{V}_A(E_T^*-E_D^*) -2bE^{*}_{D}\vec{ss} :\vec B/\sqrt{4\pi\rho}   \\
    & \qquad  =\frac{\vec{n} \cdot \nabla \rho }{4\rho }\bigg[ (E_T^*+E_C^*)(E_T^\infty+E_C^\infty)^{1/2}-(E_T^*-E_C^*)(E_T^\infty-E_C^\infty)^{1/2}  \\
    & \qquad - E_D^*\left(\sqrt{E_T^\infty+E_C^\infty}-\sqrt{E_T^\infty-E_C^\infty}\right)\bigg] -\frac{\left( E^{\infty }_{T} -E^{\infty }_{C}\right)^{1/2}\left( E^{*}_{T} +E^{*}_{C}\right)}{\lambda ^{+}_{\perp}} \\
    & \qquad +\frac{\left( E^{\infty }_{T} +E^{\infty }_{C}\right)^{1/2}\left( E^{*}_{T} -E^{*}_{C}\right)}{\lambda ^{-}_{\perp }}; \quad \label{eq:40} 
\end{split}\\ 
\begin{split}
    & \frac{\partial E^{*}_{D}}{\partial t} +\vec{U} \cdot \nabla E^{*}_{D}+\nabla \cdot \vec{U}\left[\frac{E^{*}_{D}}{2} +\left( 2b-\frac{1}{2}\right) E^{*}_{T}\right] -2bE_T^*\vec{ss}:\nabla \vec U  +2bE^{*}_{C}\vec{ss} :\nabla \vec{B}/\sqrt{4\pi\rho}  \\
    & \qquad + \nabla \cdot \vec{V}_A E_C^*  = \frac{\vec{n} \cdot \nabla \rho }{4\rho }\bigg[ E^{*}_{D}\left( \sqrt{E^{\infty }_{T} +E^{\infty }_{C}}+ \sqrt{E^{\infty }_{T} -E^{\infty }_{C}}\right) -\left( E^{*}_{T} -E^{*}_{C}\right)\left( E^{\infty }_{T} +E^{\infty }_{C}\right)^{1/2}  \\
    & \qquad  -\left( E^{*}_{T} +E^{*}_{C}\right)\left( E^{\infty }_{T} -E^{\infty }_{C}\right)^{1/2}\bigg] 
    - E^{*}_{D}\left[\frac{\left( E^{\infty }_{T} -E^{\infty }_{C}\right)^{1/2}}{\lambda^{+}_{\perp }} +\frac{\left( E^{\infty }_{T} +E^{\infty }_{C}\right)^{1/2}}{\lambda ^{-}_{\perp }} \right] ; \quad \label{eq:41} 
\end{split} \\
\begin{split}
    & \frac{\partial L^{\pm }_{*}}{\partial t} +(\vec{U} \mp \vec{V}_{A}) \cdot \nabla L^{\pm }_{*} +\frac{\nabla \cdot \vec{U}}{2} \left(L^{\pm }_{*} - \frac{L_D^*}{2} \right) +\nabla \cdot \vec{V}_A \left(\pm L^{\pm }_{*} \mp \frac{L^{*}_{D}}{2}\right)   \\
    & \qquad +bL_D^*(\nabla \cdot \vec U -\vec{ss}:\nabla \vec U \mp \vec{ss}:\nabla \vec B/\sqrt{4\pi\rho}) 
     +\frac{n\cdot \nabla \rho}{4\rho}(E_T^\infty \pm E_C^\infty)^{1/2}(L_D^*-2L_*^{\pm}) \\ 
    & \qquad = 0; \quad \label{eq:42} 
\end{split}\\
\begin{split}
    & \frac{\partial L^{*}_{D}}{\partial t} +\vec{U} \cdot \nabla L^{*}_{D} +\frac{\nabla \cdot \vec U}{2}(L_D^*-L_*^+-L_*^-) 
    +2b(\nabla \cdot \vec U -\vec{ss}:\nabla \vec U)(L_*^+ + L_*^-)\\
    & \qquad +2b\vec{ss} :\nabla \vec{B}/\sqrt{4\pi\rho}(L_*^+ - L_*^-) + \nabla \cdot \vec{V}_A(L_*^+ - L_*^-)  
     + \frac{\vec n \cdot \nabla \rho}{4\rho}\bigg[(E_T^\infty-E_C^\infty)^{1/2}(2L_*^+ - L_D^*) \\
    & \qquad + (E_T^\infty+E_C^\infty)^{1/2}(2L_*^- - L_D^*)\bigg] 
     =0, \quad \label{eq:43}
\end{split}
\end{align}
where $\vec s$ denotes the large-scale magnetic field direction.
To distinguish the structural similarity parameter $a$ for 2D incompressible turbulence from that of slab turbulence, we introduce the notation $b$.
As for 2D incompressible turbulence (subsection \ref{sec:2012}), the correlation lengths for slab turbulence are defined as 
\begin{eqnarray}
    L_{*}^{\pm} = \int \langle \vec z^{*\pm} \cdot \vec z^{*\pm '} \rangle dr =  \langle \vec z^{*\pm} \cdot \vec z^{*\pm} \rangle \lambda_{*}^{\pm} ; \label{eq:44} \\
    L_{D}^{*} = \int \langle \vec z^{*\pm} \cdot \vec z^{*\mp '} +\vec z^{*\pm '} \cdot \vec z^{*\mp}  \rangle dr = E_D^* \lambda_{D}^{*} . \label{eq:45} 
\end{eqnarray}

The transport equation for $E_T^*$ in Equation (\ref{eq:slab}) is similar to Equation (\ref{eq:2012}), thus Equation (\ref{eq:slab}) can be expressed as
\begin{align}
\begin{split}
    & \frac{\partial E_{w}^*}{\partial t} +\nabla \cdot [(\vec{U} -\vec{V}_{A} \sigma_{c}^*) E_{w}^* +\vec{U} \cdot \vec{P}_{w}^*)] \ =\vec{U} \cdot \nabla \cdot \vec{P}_{w}^*  +\frac{\vec{n} \cdot \nabla \rho }{8}\bigg[\left( E^{*}_{T} -E^{*}_{C}\right)(E_T^\infty-E_C^\infty)^{1/2} + \\
    & \quad \left( E^{*}_{T} +E^{*}_{C}\right)(E_T^\infty+E_C^\infty)^{1/2}\bigg] 
     - \frac{\rho}{2}  \bigg[ \frac{\left( E^{\infty }_{T} -E^{\infty }_{C}\right)^{1/2}\left( E^{*}_{T} +E^{*}_{C}\right)}{\lambda ^{+}_{\perp }} -\frac{\left( E^{\infty }_{T} +E^{\infty }_{C}\right)^{1/2}\left( E^{*}_{T} -E^{*}_{C}\right)}{\lambda ^{-}_{\perp }}\bigg]  \quad \label{eq:46}  
\end{split}
\end{align}
Here, the turbulence pressure tensor is defined as 
\begin{align}
    \vec{P}_{w}^{*} & \equiv  \left[ \frac{E_{w}^{*}}{2} +\left( 2b-\frac{1}{2}\right) \sigma_{D}^* E_{w}^{*}\right]\vec{I} -2b\sigma_D^*E_w^*\vec{n}\vec{n} \nonumber \\
    &= \left[b\rho \langle \vec{u_1}^2 \rangle + (1-2b)\left (\frac{\langle \vec {b^*}^2 \rangle}{8\pi} \right)\right] \vec I - b\left( \rho \langle \vec{u}^2_1 \rangle - \frac{\langle {\vec{b}^*}^2 \rangle}{4\pi} \right)\vec{n}\vec{n} ,\label{eq:47}
\end{align}
where $\vec u_1$, $\vec b^*$ are the fluctuating velocity and magnetic field for slab turbulence, respectively.
It is worth noting that if $\sigma_D\neq 0$ and $b>1/2$, it is possible for
the isotropic part of the turbulence pressure tensor to become negative.
To avoid these unphysical situations, it is necessary that care be exercised in choosing the value of the structural similarity parameter.

We illustrate different forms of the turbulence pressure tensor under different assumptions. In the case that $\sigma_D^*=0$, the slab turbulence pressure tensor, Equation \ref{eq:47}, is given by,
\begin{equation}
    \vec P_w^* = \frac{E_w^*}{2}\vec I= \frac{\langle \vec {b^*}^2 \rangle}{8\pi} \vec {nn}, \label{eq:48}
\end{equation}
which is isotropic and resembles the pressure exerted by Alfv\'en waves. 
For the case $b=1/2$, then the slab turbulence pressure tensor can be expressed as
\begin{equation}
    \vec P_w^* = \left[ \frac{E_w^*}{2} + \sigma_D^*E_w^* \right]\vec I -\sigma_D^* E_w^* \vec{nn} 
     = \frac{\rho \langle \vec u_1^2 \rangle}{2} \vec I -\left(\frac{\rho\langle \vec u_1^2\rangle}{2}-\frac{\langle \vec {b^*}^2 \rangle}{8\pi}\right)\vec{nn} \label{eq:49}
\end{equation}
If instead we assume $b=0$, the slab turbulence pressure tensor $\vec P_w^*$ is
\begin{equation}
    \vec P_w^* = E_w^* \left (\frac{1}{2} - \frac{\sigma_D^*}{2} \right) \vec I = \frac{\langle \vec {b^*}^2 \rangle}{8\pi} \vec I. \label{eq:50} 
\end{equation}

\section{Conclusions}\label{sec:conclusion}
In a formal sense, when considering the relationship of the ideal incompressible MHD equations to the ideal compressible MHD equations that results in the theory of nearly incompressible MHD \citep{Zank_Matthaeus_1993}, essentially the two limits of large plasma beta and small or order one plasma beta are relevant. Based on this ordering, 
\citet{Zank_etal_2012} (large beta) and \citet{Zank_etal_2017} (beta order 1) derived two sets of equations describing 
 the evolution of incompressible and nearly incompressible MHD turbulence in inhomogeneous flows as expressed through ``moments'' of the fluctuating Els\"asser variables.  The large beta limit yields a transport formalism that is at leading order based on the fully 3D incompressible MHD equations whereas the beta order 1 or $\ll 1$ limit yields a superposition of quasi-2D  MHD as the leading-order or dominant component and a minority slab component. Neither the \citet{Zank_etal_2012} nor
 the \citet{Zank_etal_2017} derived a conservation form of the transport equations for the energy in the fluctuations. Expressing the fluctuating energy in the form of a conservation law is an important check on the physical and mathematical consistency of the turbulence transport formalism. Despite the evident complexity of the underlying turbulence transport equations, we show here that in both limits, 
the transport equations for the turbulence energy can be expressed in conservative form through the introduction of generalized forms of the pressure tensor for the fluctuating velocity and magnetic field components.  

Our principle results are Equations (\ref{eq:29}) (beta $\gg 1$), (\ref{eq:37}) (quasi-2D turbulence, beta $\sim 1$),  and (\ref{eq:46}) (slab turbulence, beta $\sim 1$) together with the respective definitions of the turbulence pressure (\ref{eq:23}), (\ref{eq:38}), and (\ref{eq:47}). The fluctuating energy conservation laws for the beta $\gg 1$ and the slab turbulence beta order $\sim 1$ cases resemble formally the well-known WKB transport
equations
for the energy density of linear Alfv\'en waves in an inhomogeneous flow. However, the analogy is not close for multiple reasons; i) the turbulence transport formalism does not assume that the fluctuations are small-amplitude or linear; ii) the turbulence pressure tensor $\vec{P}_w$ is significantly different from the wave pressure tensor of WKB models, containing typically both the energy densities of the fluctuating velocity and magnetic fields. The relevant anisotropies of the underlying
turbulence are also contained in the turbulence pressure tensors as expressed through the structural similarity parameters that represent a closure relation between the trace and the covariance terms in the one-point Els\"asser energy tensor terms; iii) The energy density flux in the turbulence conservation laws is similar to the WKB formalism in that the turbulence form contains the cross helicity, although in the turbulence case, the cross helicity is governed by an independent
turbulence transport equation that must be solved in conjunction with the energy transport equation; iv) Finally, the role of dissipation is properly incorporated in the conservation laws and is based on a Kolmogorov formalism (equally the dissipation terms can be treated using an Iroshnikov-Kraichnan formalism \citep{Ng_etal_2010} provided one is not modeling quasi-2D turbulence \citep{Zank_etal_2020}). The conservation form of the dominant quasi-2D turbulence case (beta $\sim 1$), Equation (\ref{eq:38}), resembles too the WKB formalism in some terms but is quite different in that the Alfv\'en velocity term is absent entirely. This of course is because the fluctuations are quasi-2D structures such as flux ropes and not Alfv\'en waves. The differences i) - iv) above also apply to the quasi-2D turbulence conservation law. For the beta $\sim 1$ case, the separation into dominant and minority components means that the quasi-2D turbulence energy density $E_w^{\infty}$ and pressure $\vec{P}_w^{\infty}$ and the slab turbulence energy density $E_w^*$ and pressure $\vec{P}_w^*$ can be combined to obtain the total turbulence energy density and pressure tensor contribution.

The conservation forms of the turbulence energy density equations cannot be solved in isolation since they are coupled to the evolution of the cross helicity, the residual energy, and the various correlation lengths. Nonetheless, the conservation form can be used in place of the total energy transport equation formalism used in \citet{Zank_etal_2012} and \citet{Zank_etal_2017} and applications thereof. If one chooses to impose certain constraints on e.g., the cross helicity or residual energy \citep{Zank_etal_1996, Zank_etal_2012}, the conservation form is particularly useful in deriving simplified forms of the turbulence transport equation, many of which are readily amenable to analytic solution \citep{Zank_etal_1996}. 

In conclusion, we have derived the conservative form of the turbulence energy transport equations, and shown explicitly the dissipation terms and derived generalized forms of the turbulence pressure tensor. We anticipate that our results will be useful for a range of important and interesting problems in solar, stellar, and other large-scale astrophysical winds, cosmic ray physics, shock waves, and especially the heating of the solar corona and solar wind.

\section*{Acknowledgement}
We acknowledge the partial support of an NSF EPSCoR RII-Track-1 Cooperative Agreement OIA-1655280, partial support from a NASA Parker Solar Probe contract SV4-84017, partial support from a NASA LWS grant 80NSSC20K1783, and partial support from a NASA IMAP subaward
under NASA contract 80GSFC19C0027. G.P. Zank was partially supported by an NSF CDS\&E Award 2009871. 

\bibliography{paper}
\bibliographystyle{aasjournal}
\end{document}